\documentstyle[preprint,aps,psfig]{revtex}
 
\newcommand {\be}{\begin{equation}}
\newcommand {\ee}{\end{equation}}
\newcommand {\bey}{\begin{eqnarray}}
\newcommand {\eey}{\end{eqnarray}}

\begin{document}
\draft

\title{Entropy potential and Lyapunov exponents}

\author{Stefano Lepri$^{1\dagger}$, Antonio Politi$^{2\ddagger}$ 
and Alessandro Torcini$^{3\dagger}$}

\address{
$^1$ {\it Max-Planck-Institut f\"ur Physik Komplexer Systeme
D-01187 Dresden, Germany} \\
$^2$ {\it Istituto Nazionale di Ottica I-50125 Firenze, Italy}\\
$^3$ {\it Dipartimento di Energetica ``S. Stecco'' I-50139 Firenze, Italy}}

\date{\today}
\maketitle
\begin{abstract}
According to a previous conjecture, spatial and temporal Lyapunov exponents
of chaotic extended systems can be obtained from derivatives of a suitable
function: the entropy potential. The validity and the consequences of this
hypothesis are explored in detail. The numerical investigation of a
continuous-time model provides a further confirmation to the existence of
the entropy potential. Furthermore, it is shown that the knowledge of the
entropy potential allows determining also Lyapunov spectra in general
reference frames where the time-like and space-like axes point along generic
directions in the space-time plane. Finally, the existence of an entropy
potential implies that the integrated density of positive exponents
(Kolmogorov-Sinai entropy) is independent of the chosen reference frame.
\end{abstract}
\vskip .5cm

\centerline{\bf Lead paragraph}
\vskip 1.cm
Out-of-equilibrium systems give rise to a rich variety of phenomena which
are often dealt with {\it ad hoc} methods. Perhaps the main difficulty lying
behind the development of general tools is the lack of an Ansatz equivalent
to the Gibbs-Boltzmann weight, which allows estimating a priori the
probability of each configuration of an equilibrium system. Although the
problem cannot be (easily) overcome, it is tempting to explore the
possibility to extend and apply concepts like Lyapunov exponents to extract
information about the invariant measure of a high-dimensional system.
For instance, on the one hand it is well known that, through the Kaplan-Yorke
formula, it is possible to estimate the fractal dimension of a finite (low)
dimensional attractor; on the other hand, however, it is unclear how the
local dynamics arising in different regions of an extended system combine
together to determine the global invariant measure. It is our opinion that
a sensible answer will be given only after having extended the Lyapunov
analysis, to account for spatial propagation as well as temporal
instabilities. The chronotopic approach, introduced in
Ref.~\cite{lyap1,lyap2} represents a tentative construction that goes in
this direction, introducing a sort of dispersion relations in chaotic
systems.  In this paper, we test the general validity of the method
both with numerical simulations and by investigating its internal coherence.
\newpage

\section{Introduction}

Lyapunov exponents, providing information on the evolution in tangent space
are very useful in that they allow characterizing the invariant measure
\cite{eck}. This is particularly clear in the case of hyperbolic systems as 
revealed by the construction of the Bowen-Ruelle-Sinai measure \cite{brs}, 
but is also effectively true for the general and more realistic class of
smooth non-hyperbolic systems. However, an effective exploitation of these
ideas in spatially extended systems is hindered by the
infinite-dimensionality of the phase-space and by the existence of
propagation phenomena that requires a more detailed understanding besides
that one provided by the knowledge of the usual Lyapunov spectrum
\cite{cross}. In order to reach a more complete comprehension
of phenomena occurring in tangent space, two families of Lyapunov spectra
characterizing temporal, resp. spatial, dynamics of infinitesimal
perturbations have been introduced and discussed \cite{lyap1}.
Subsequently \cite{lyap2}, it has been shown that the two families are
not independent of one another. More precisely, it has been conjectured that
all the information on Lyapunov exponents can be obtained from a single
observable: the {\it entropy potential} $\Phi$, a function of the temporal
and the spatial growth rates $\lambda$, $\mu$, respectively. The name
``entropy potential'' follows from the observation that 
$\Phi(\lambda=0,\mu=0)$
coincides with the density of Kolmogorov-Sinai entropy \cite{grass}.
However, such a conjecture is based only on a few simple examples that can
be analytically worked out and on numerical simulations performed for a
chain of coupled maps. One of the aims of the present paper is to
strengthen the validity of the conjecture by performing new tests in a more
realistic system.

Moreover, we intend to show a further connection between spatial and temporal
Lyapunov exponents by studying the evolution of perturbations along generic 
``world-lines'' in the space-time, i.e. along directions other than the 
natural space and time axes. The extension of the usual definition of 
Lyapunov exponents to this more general class of frames, partially discussed 
in \cite{bidime}, is rather appropriate for characterizing patterns with 
some anisotropy. Here, we prove that this seemingly more general class of 
{\it spatiotemporal} exponents can be derived from the knowledge of spatial 
and temporal Lyapunov spectra, which thus confirm to contain all the 
relevant information. 

The present paper is organized as follows. In Section II we recall 
the definition of the entropy potential and present some
new test for its existence. Sec. III is devoted to Lyapunov 
analysis in tilted reference frames i.e. to the definition of spatiotemporal
exponents, while Sec. IV deals with their relationships with more standard
dynamical indicators. Some conclusive remarks are finally reported in
Sec.~V.

\section{Lyapunov spectra from an Entropy potential}

In this Section, we first introduce the main classes of models 
employed in the investigation of spatio-temporal chaos. Then, we
recall the notion of temporal (TLS) and spatial (SLS) Lyapunov spectra 
with a particular emphasis to their connection with the entropy potential. 
The very existence of the latter is then investigated in the last part,
for a specific continuous-time model.

Reaction-diffusion systems are among the most relevant models for the study
of spatiotemporal chaos. The standard form for the evolution equations is 
\cite{cross}
\be
\label{pde}
\partial_t {\bf u} = {\bf F}({\bf u}) + {\bf D} \partial_x^2 {\bf u} \quad ,
\ee
where the field ${\bf u}(x,t)$ is defined on the domain $[0,L]$, with
periodic boundary conditions ${\bf u}(0,t)={\bf u}(L,t)$. The nonlinear 
function ${\bf F}$ accounts for the local reaction dynamics, while the 
matrix ${\bf D}$ represents the spatial coupling induced by diffusion. 

A simplified class of models can be obtained by spatial discretization,
i.e. by considering a 1D lattice of coupled oscillators 
\cite{aranson}
\be
\label{ode}
\dot {\bf u}_i = {\bf F}({\bf u}_i) + D\left( {\bf u}_{i+1}-2 
{\bf u}_i +{\bf u}_{i-1}\right) \quad ,
\ee
where the index $i$ labels each site of a lattice of length $L$ 
(assuming again periodic boundary conditions). 

A further simplification
is achieved by discretizing also the time, i.e by considering coupled 
map lattices (CML) \cite{kaneko}, that in their usual form read as
\be
\label{mappa}
  u^i_{n+1} = f\left(
      {\varepsilon \over 2}u^{i+1}_n+
      (1-\varepsilon)u^i_n+ 
      {\varepsilon \over 2}u^{i-1}_n
     \right)\quad ,
\ee
where $n$ is the time index, and $\varepsilon$ gauges the diffusion 
strength. The function $f$, mapping a given interval $I$ of the real 
axis onto itself, simulates the nonlinear reaction process. In 
particular, we will focus on the homogeneous $f(x)=rx \pmod{1}$
and logistic $f(x)=4x(1-x)$ CML with $u_n^i\in[0,1]$.

CML models have been also proposed to mimic 1D open-flow systems 
\cite{como}, characterized by flux terms in the field equation. 
For example the model
\be
\label{amappa}
  u^i_{n+1} = f\left(
      \varepsilon (1-\alpha) u^{i-1}_n+
      (1-\varepsilon)u^i_n+
      \varepsilon\alpha u^{i+1}_n \right)\quad,
\ee
with the parameter $\alpha$ bounded between 0 and 1, accounts for the 
possibility of an asymmetric coupling, corresponding to first order 
derivatives in the continuum limit. 

Let us denote with $\delta(x,t)$ a generic perturbation that can be assumed
to possess an exponential profile both in space and time,
\be
   \delta(x,t) =  a(x,t) \exp (-\mu x + \lambda t) \quad .
\ee
Depending whether $\mu$, or $\lambda$ is considered as a free parameter, one
can define either the temporal or the spatial Lyapunov spectrum. Let us
first discuss the temporal Lyapunov spectrum $n_\lambda(\lambda,\mu)$:
it is obtained in the usual way by following the evolution in tangent space of
a perturbation of the form $b(x,t) e^{\mu x}$, where $\mu$ is
fixed a priori. Obviously, for $\mu = 0$, one recovers the standard Lyapunov
spectrum. Alternatively, one can consider the temporal growth rate $\lambda$ 
as a free parameter and thereby determine the spatial spectrum $n_\mu(\lambda,\mu)$. 
The latter procedure is symmetric to the previous one: $\mu$ and $\lambda$ are
mutually exchanged, as well as $x$ and $t$, i.e. the tangent dynamics is
followed along the spatial direction.

As a result of the above two approaches one is confronted with a set of four
variables: $\mu$, $\lambda$, $n_\mu$, and $n_\lambda$. The main conclusion of
\cite{lyap1} is that any two of them can be taken as independent variables,
the other two variables providing a complete characterization of
spatio-temporal stability properties. Furthermore, in \cite{lyap2}, we have 
conjectured that both integrated densities are not actually independent of one another, 
but can be obtained from a single function $\Phi(\lambda,\mu)$,
\bey
\label{enpot2}
  &&\partial_\lambda \Phi  = n_\lambda \\
  &&\partial_\mu \Phi  = n_\mu \nonumber \quad ,
\eey
that is called entropy potential as it coincides with the usual
Kolmogorov-Sinai entropy density along a suitable line (see Sec. IV).

Since equivalent descriptions are obtained by choosing any pair of
independent variables, the appropriate entropy potential is obtained by
means of a Legendre transform in the new variables.
For instance, if one wishes to consider $\lambda$ and $n_\lambda$ as new
conjugate variables, the new entropy potential $\Psi(n_\lambda,\lambda)$ is
given by
\be
  \Psi \equiv \lambda n_\lambda - \Phi \; ,
\ee
and the following relations hold
\bey
\label{enpot1}
  &&{ \partial \Psi \over \partial n_\lambda} = \lambda \\
  &&{ \partial \Psi \over \partial \mu} = -n_\mu \quad .\nonumber
\eey
The existence of a potential has been conjectured on the basis of analyticity
properties of the relationship linking $\mu$ and $\lambda$.
The key steps to prove, e.g. (\ref{enpot1}), are (i) the introduction of
complex spatial and temporal ``growth rates'' $\tilde \mu$ and 
$\tilde \lambda$, respectively; (ii) the identification (apart from
a sign and a multiplicative factor) of $Im(\tilde \lambda)$ and
$Im(\tilde \mu)$ with $n_\mu$ and $n_\lambda$, respectively (obviously,
$Re(\tilde \lambda) = \lambda$ and $Re(\tilde \mu) = \mu$).

As a matter of fact, we have been able to substantiate the second point only
in simple cases where the linear stability analysis reduces to the an
eigenvalue equation for a fixed point or a periodic orbit. In the fully
aperiodic regime, there is only numerical evidence in a chain of coupled
logistic maps, where a test on the existence of a potential has been
performed by computing a certain circulation integral \cite{lyap2}.

In order to test the general validity of our main Ansatz, we consider
now a continuous-time model. Since the generation mechanisms of the
multipliers in tangent space are not important for our conclusions,
we have assumed that they are the result of stochastic processes, thus
escaping the need to integrate also some nonlinear equations in
phase-space. We shall refer to a model of coupled oscillators (see
Eq.~\ref{ode}), whose evolution in tangent space can be written as
\be
\dot{\delta u}_i = m_i(t) \delta u_i + D(\delta u_{i+1}+\delta u_{i-1})
\label{rndosci}
\ee
where the random multipliers $m_i(t)$ are indirectly defined through their
power spectrum
\be
S(\omega) = S_0 \left[\exp\left(-{(\omega-\omega_0)^2\over \sigma}\right)
+\exp\left(-{(\omega+\omega_0)^2\over \sigma}\right)\right] \quad .
\label{powspe}
\ee
This choice implies that $m_i(t)$ is an analytic function of time and
possesses some degree of periodicity as one expects to be the case of many
deterministic chaotic signals.

 From a numerical point of view, we have decided to compute the derivative
appearing in (\ref{rndosci}) through a simple finite difference
(Euler) scheme. This approach, although not the most refined one,
allows treating in a consistent way the temporal integration and the
recursive iteration in space. Indeed, by expressing the perturbation as
$\delta u_i(t) = v_i(t) \exp{(-\mu i +\lambda t)}$, 
Eq.~(\ref{rndosci}) can be rewritten as
\be
v_i(t+ \Delta t) {\rm e}^{\lambda \Delta t} = v_i(t) + \Delta t
\left[ m_i(t) v_i + D(v_{i+1} {\rm e}^{\mu}+v_{i-1} 
{\rm e}^{-\mu}) \right] \quad ,
\label{discret}
\ee
which can be easily iterated either in space or time to determine the
TSL and the SLS. In principle, one should choose a sufficiently small
integration time step; in practice the spatial recursion becomes soon
unfeasible, since the dimension of the ``temporal'' phase space is
proportional to $1/\Delta t$. Accordingly, we have considered a few
different values of $\Delta t$, namely 0.2, 0.1, and 0.05, paying more
attention to testing the existence of the potential in each case, 
rather than to computing the actual limit spectra (i.e. to the limit
$\Delta t\to 0$). In fact, although for a finite value of $\Delta t$,
one cannot assert, rigorously speaking, to have simulated Eq.~(\ref{ode}),
model (\ref{discret}) can nevertheless be considered as a dynamical system
in itself worth being studied (actually, it is a CML with a different
coupling scheme from that usually assumed).

\begin{figure}
\psfig{figure=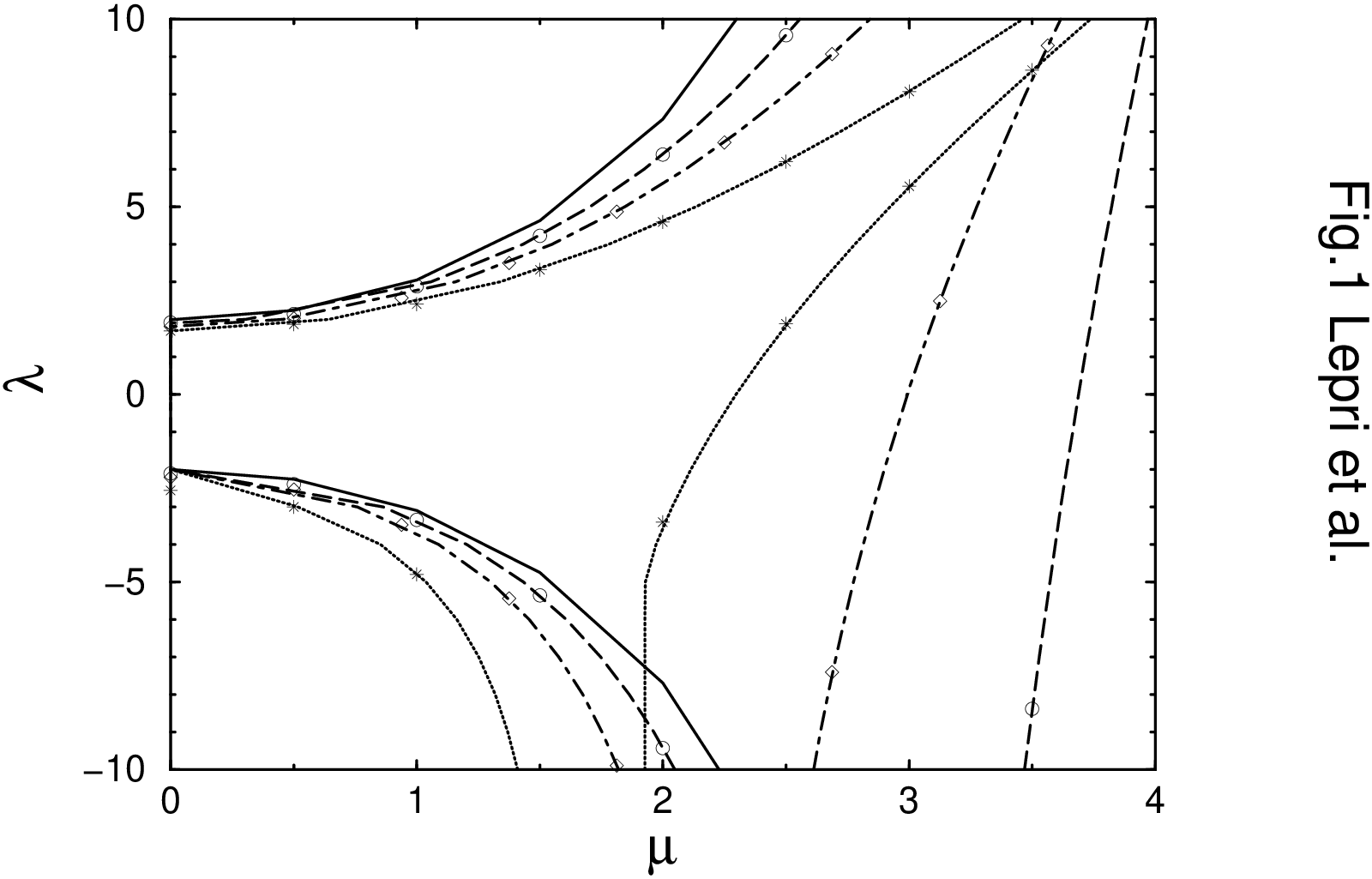,angle=-90,height=9truecm,width=12truecm}
\caption{Boundaries $\partial{\cal D}$ for the 
model (\protect{\ref{rndosci}}) as obtained from TLS (lines)
and SLS (symbols) for three different values of the
time step $\Delta t$ : 0.2 (dotted line and asterisks), 
0.1 (dot-dashed line and diamonds) and 0.05 (dashed line
and circles). The asymprotic boundaries are also reported
for $\Delta t = 0.00625$ (solid line).
}
\label{conti}
\end{figure}

In Fig.~\ref{conti}, we have reported the borders of the spectra in the
$(\mu,\lambda)$-plane as estimated from the SLS and the TLS. 
The agreement between the two sets of lines for every value of $\Delta t$
provides a first indication that each approximation of the continuous model
is in itself consistent with the findings of \cite{lyap1}. Each border is
made of 4 distinct lines: an upper and a lower curve representing the set of
maxima and minima of TLS; two symmetric lines representing the extrema of
the SLS. For $\Delta t\to 0$, we expect that the first two curves converge
to an asymptotic shape, while the latter two diverge to $\pm \infty$. In
fact, the maximum and the minimum spatial divergence rate are limited by the
time step $\Delta t$. This is a general feature of continuous time systems
as already remarked in \cite{lyap1}.

The variation of the lateral borders with $\Delta t$ indeed confirms our
expectations, while the dependence of the upper and lower curves
(especially at high $|\mu|$-values) indicates, instead, that the convergence
to the asymptotic form is rather slow: this is clearly confirmed by the
comparison with the border of the TLS determined for a much smaller value of
$\Delta t$ (0.00625) (see solid curve in Fig.~\ref{conti}).

In any case, the most relevant test that we wanted to perform concerns the
the existence of an entropy potential $\Phi$. This can be numerically done
by computing the integral
\be
\int_A^B \, (n_\mu,n_\lambda)\cdot \vec w \; ds
\ee
along different paths joining generic pairs $A$, $B$ of points in the
$(\mu,\lambda)$ plane. If the integral is independent of the path,
we have a clear indication of the existence of a potential. All tests that
we have performed for different choices of points and of the
time-integration steps have been successfull, yielding results that are
equal within the numerical accuracy. In Table I, we report a subset of
the results, all referring to the same end points but different values
of $\Delta t$. We consider this as a further clear evidence that the
potential entropy indeed exists and is a rather general feature of
at least 1D spatio-temporal systems.

\section{Spatiotemporal exponents}

In the perspective of a complete characterization of space-time chaos, one 
should consider the possibility of viewing a generic pattern as being
generated along directions other than time and space axes. In fact, once a
pattern is given, any direction can, a priori, be considered as an
appropriate ``time'' axis. Accordingly, questions can be addressed about the
statistical properties of the pattern when viewed in that way.
The extension of the Lyapunov formalism to generic orientations of the 
space-time coordinates does not follow simply from an abstract need of
completeness; it also results from the attempt of generalizing the nonlinear
time-series analysis to patterns. The existence of low-dimensional chaos led
many researchers to investigate the possibility whether a given irregular 
temporal signal might be the consequence of a few nonlinearly coupled degrees
of freeedom. The existence of space-time chaos leads to the
equivalent question whether a given pattern is the result of a deterministic
1D nonlinear process. At variance with temporal signals which can be generated
only by moving forward or backward in time, in the case of patterns, the
identification of the most appropriate spatial and temporal directions is
a new and unavoidable element of the game. For this reason, in the next
subsection we shall introduce the notion of spatiotemporal exponents.

\subsection{Definitions}

For the moment, we assume that the pattern is continuous both along space
and time directions; we shall discuss later how the definitions can be extended
to CML models. Let us consider a given spatiotemporal configuration of the 
field ${\bf u}(x,t)$, generated, say, by integrating Eq.~(\ref{pde}). 
When arbitrary directions are considered in the $(x,t)$ plane, the coordinates
must be properly scaled in order to force them to have the same dimension. We 
choose to multiply the time variable by $c$, where $c$ is a suitable constant 
with the dimension of a velocity.  Moreover, let $\vartheta$ denote the 
rotation angle of the tilted frame $(x',ct')$ with respect to the initial one 
$(x,ct)$, adopting the convention that positive angles correspond to clockwise 
rotations. Sometimes it will be more convenient to identify the new frame by
referring to the velocity $v = c\tan\vartheta$. The limit cases $v=0$ 
($\vartheta=0$) and $v=+\infty$ ($\vartheta=\pi/2$) correspond to purely
temporal and purely spatial propagations, respectively. 
The coordinate transformation reads as
\bey \label{trasfo}
&&ct' = \beta \left(ct +{v\over c} x \right) \\ 
&&x' = \beta \left( -vt + x \right) \quad , \nonumber
\eey 
where $\beta \equiv 1/\sqrt{(1+v^2/c^2)}$. The physical meaning of $v$ is
transparent: it can be interpreted as the velocity in the old 
frame of a point stationary in the tilted frame (constant $x'$).

The new field ${\bf u}(x',ct')$ can be thought of as being the result of the
integration of the model derived from the original one after the change of 
variables (\ref{trasfo}). Although it is not obvious whether the invariant 
measure in the initial frame is still attracting in the new frame 
(see Ref.~\cite{bidime} for a discussion of this point), one can anyhow study 
the stability properties by linearizing and defining the Lyapunov exponents in 
the usual way. 

In CML models, the discreteness of both the space and the time lattice 
leads to some difficulties in the practical construction of tilted frames. In 
fact, only rational values of the velocity $v$ can be realized in finite
lattices (in this case, it is natural to assume that the lattice spacing is 
the ``same'' along the spatial and the temporal directions and, accordingly,
to set $c=1$). Moreover, writing the explicit expression of the model 
requires introducing different site types. For this reason, we discuss in the 
following the simplest nontrivial case $v=1/2$, the generalization to other
rational velocities being conceptually straightforward.

\begin{figure}
\psfig{figure=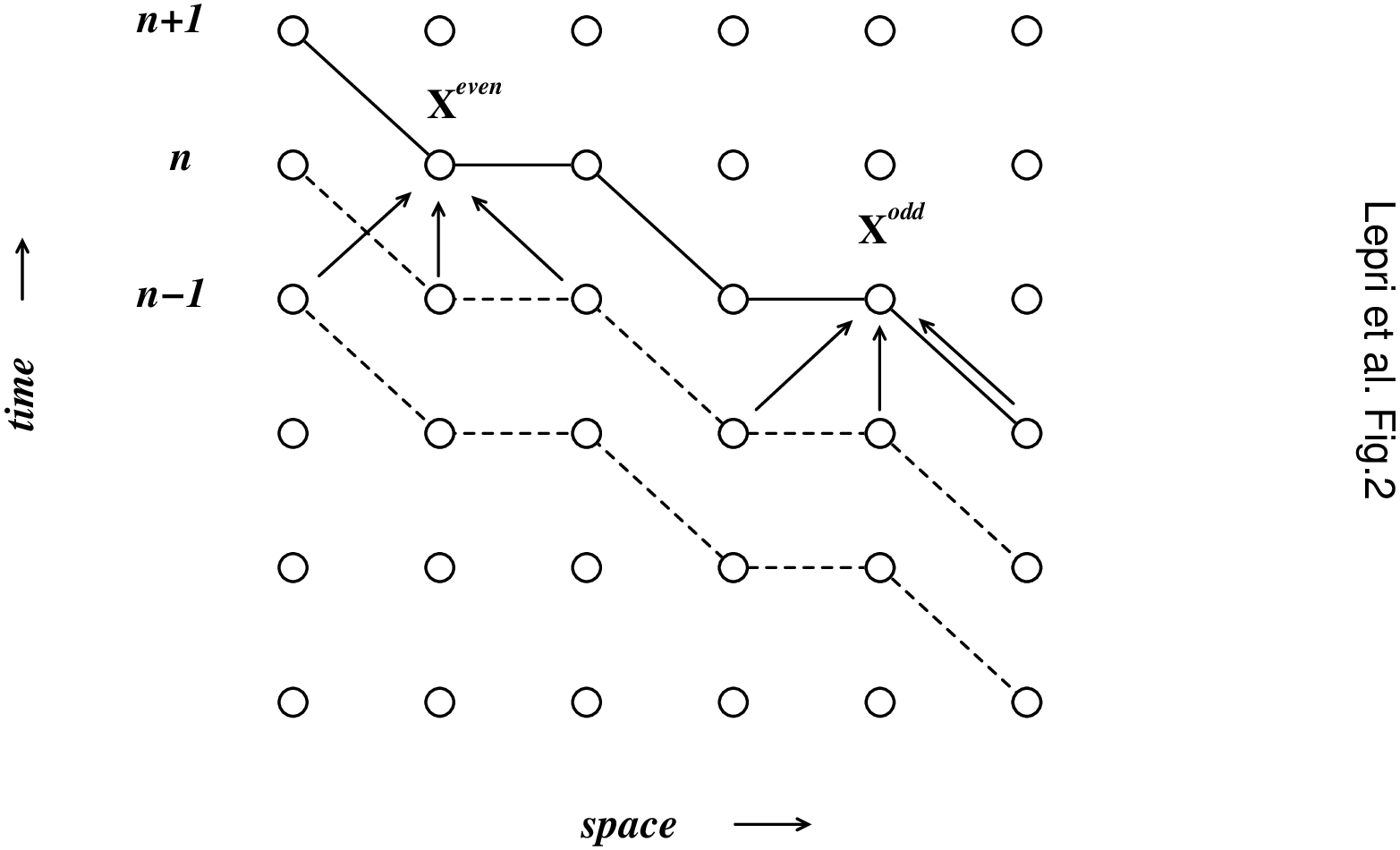,angle=-90,height=9truecm,width=12truecm}
\caption
{Lattice implementation of the definition of 
spatiotemporal Lyapunov exponents for $v=1/2$.}
\label{tilt}
\end{figure}

A generic spatial configuration in the tilted frame is defined by sites of the 
spatiotemporal lattice $(i,n)$ connected by alternating horizontal (as in the 
usual case) and diagonal bonds (see Fig.~\ref{tilt}). By suitably adjusting 
the relative fraction of the two types of links, all rotations between 0 and
$\pi/4$ can be reproduced.
The explicit expression of the updating rule requires a proper numbering of 
the consecutive sites. Moreover, as seen in Fig.~\ref{tilt}, it involves the 
``memory'' of two previous states. 

Finally, an exact implementation of the mapping rule requires acausal boundary
conditions, since the knowledge of future (in the original frame) states is
required \cite{bidime} (this is a general problem occurring also in the 
continuous case). As we are interested in the thermodynamic limit, we bypass
the problem by choosing periodic boundary conditions. Such a choice has been 
shown not to affect the bulk properties of the dynamical evolution 
\cite{bidime}.\par
In the updating procedure, two different cases are recognized: the variable
$u$ is either determined from the past values in the neighbouring sites, or
it requires the newly updated $u$-value on the right neighbour (see
Fig.~\ref{tilt}). For $v=1/2$, this can be done by simply distinguishing
between even and odd sites,
\begin{eqnarray}
\label{v2}
&&X^{2i}_{n+1} 
   = f\left(
      {\varepsilon \over 2} Y^{2i-1}_n+
      (1-\varepsilon)X^{2i}_n+ 
      {\varepsilon \over 2} X^{2i+1}_n \right)\\ 
&&X^{2i+1}_{n+1} 
   = f\left(
       {\varepsilon \over 2} X^{2i}_n +
       (1-\varepsilon)X^{2i+1}_n+ 
       {\varepsilon \over 2}
       X^{2i+2}_{n+1}\right)\quad ,\nonumber 
\end {eqnarray}
where $i=1,\ldots,L/2$ ($L$ is assumed to be even for simplicity), while
\be
Y^j_{n+1} \equiv X^j_n \quad ,
\ee
are additional variables introduced to account for the dependence at time
$n-1$. Taking into account that $X_{n+1}^{2i+2}$ can be determined from
the $X$ and $Y$ variables at time $n$, the mapping can be finally expressed
in the usual synchronous form $(X_n^i,Y_n^i)\to (X_{n+1}^i,Y_{n+1}^i)$, but
with an asymmetric spatial coupling with next and next-to-next nearest
neighbours. The Lyapunov exponents $\eta_j$ can now be computed with the 
usual technique \cite{benettin}.

In analogy with the original model, we expect again that, in the limit of 
infinitely extended systems, the set of exponents $\eta_j(v)$ will converge 
to an asymptotic form,
\be
 \eta_j(v) \to \eta(v,n_\eta)\quad, 
\ee
where $n_\eta=j/L$ is the corresponding integrated density. We will refer to 
this function as the spatiotemporal Lyapunov spectrum (STLS). In the limit 
cases $v=0,+\infty$ ($\vartheta=0,\pi/2$), the STLS reduces to the 
standard temporal and spatial spectrum, respectively.

The recursive scheme (\ref{v2}) implies an increase of the phase-space 
dimension by a factor (1+1/2) (in general $1+v$). Actually, as
we will argue, these new degrees of freedom are not physically relevant.
However, for consistency reasons with the original rescaling of the spatial 
variable, we choose to normalize the spatiotemporal density between 0 and 
$1+v$ (the time units are, instead, left unchanged by the above construction).

The generalization to asymmetric maps (\ref{amappa}) is straightforward: it
removes the degeneracy $v \to -v$. Numerical results for logistic maps,
indicate that the dependence of the positive exponents on the velocity is
quite weak in the fully symmetric case $\alpha =1/2$ (for instance, the
maximum exponent exhibits a 20\% variation in the whole $v$ range), while
it is remarkable for asymmetric couplings. In every case, the negative part
of the spectrum sharply changes with the velocity. This is consistent with
the results obtained for delayed maps in Ref.~\cite{bidime}.

\subsection{Representation in the $(\mu,\lambda)$ plane}

Spatiotemporal exponents can be put in relation with $\mu$ and $\lambda$ 
by rewriting the general expression for a perturbation in a frame rotated by
an angle $\vartheta$,
\be
\label{rota}
   \exp(\mu x+\lambda t)=\exp(\mu'x'+\lambda' t')\quad .
\ee
Such an equation induces a rotation of the same angle in the
$(c\mu,\lambda)$ variables,
\bey
\label{trasfolamu}
&&\lambda' = \beta \left( \lambda +v \mu \right) \\ 
&&\mu' = \beta \left( -(v/c^2)\lambda + \mu \right) \quad . \nonumber
\eey 
The above equations allow studying the stability with respect to generic
perturbations with an exponential profile along $x'$. For simplicity, we 
shall consider only uniform perturbations, 
\be
\label{eta}
   \exp(\mu x+\lambda t)=\exp(\eta t') \quad ,
\ee
where the growth rate $\eta$ denotes the spatiotemporal exponent. Notice that
we have changed notations from $\lambda'$ to $\eta$, to understand that the 
condition $\mu' = 0$ is fulfilled. From the second of Eq.~(\ref{trasfolamu}),
uniform perturbations in the rotated frame correspond to points along the line
$\cal L$ 
\be
\label{velvel}
\lambda = c^2\mu/v  \quad ,
\ee
in the $(\mu,\lambda)$ plane.

Whenever the evolution of an exponentially localized perturbation of type
(\ref{rota}) is considered, it is natural to introduce the quantity
$\hat V=\lambda/\mu$, which can be interpreted as the velocity of the front
\cite{disturbi}. Eq.~(\ref{velvel}) connects this velocity with that of the
rotated frame, 
\be
\hat V =c^2/v \quad .
\ee
Therefore, on the basis of definition (\ref{eta}), $(x',ct')$ can be
interpreted as the reference frame in which the front associated with the
perturbation propagates with an ``infinite'' velocity.

\begin{figure}
\psfig{figure=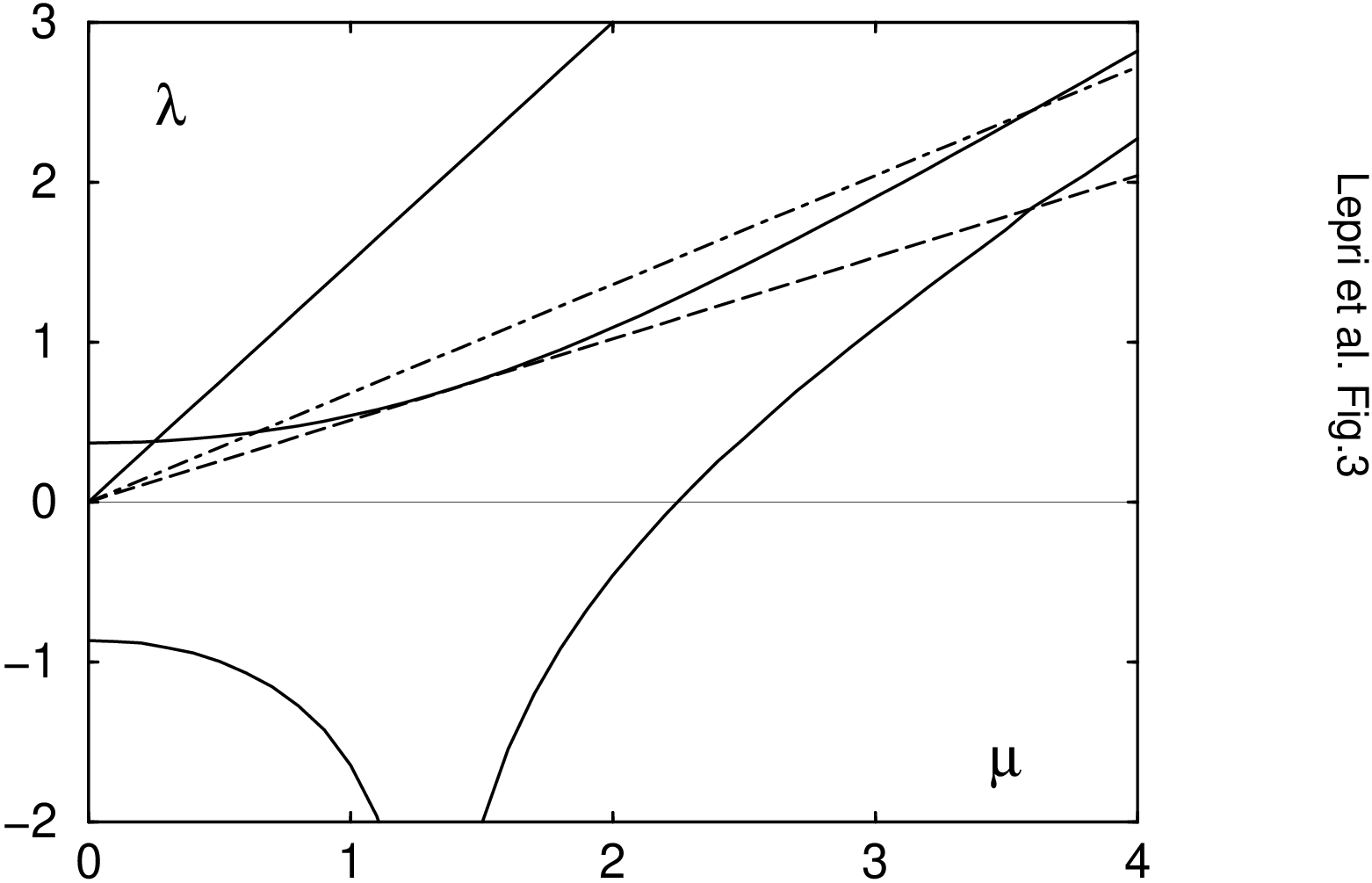,angle=-90,height=9.5truecm,width=14truecm}
\caption{Plot of the boundary $\partial{\cal D}$ and of the line
$\lambda= v \mu$ in the three velocity regimes for the logistic
CML $\varepsilon=1/3$. The three lines refer to the different 
cases $v < 1$ (solid), $ 1 < v < 1/V_*$ (dot-dashed) and 
$v=1/V_*$ (dashed).}
\label{bordi}
\end{figure}

The explicit expression for $\eta$ is
\be
\label{distan}
\eta = {\rm sign}(\lambda) \sqrt{\lambda^2 + (c \mu)^2} = \lambda/\beta \quad .
\ee
Such a relation can be turned into a self-consistent equation for the
maximum Lyapunov exponent $\eta_{\rm max}$ by imposing the constraint that
the pair $(\mu,\lambda)$ lies on the upper border 
$\lambda = \lambda_{\rm max}(\mu)$ of the spectrum of the Lyapunov exponents,
namely 
\be
\label{selfcon}
 \eta_{\rm max} = {1 \over \beta} \lambda_{\rm max} \left(\mu={v\over c^2} 
   \beta\eta_{\rm max} \right)\quad .
\ee 
Some ambiguities arise when velocities $v>c$ are considered, since the line
$\cal L$ intersects $\lambda_{\rm max}(\mu)$ in two points as seen in
Fig.~\ref{bordi}. This phenomenon was already noticed in Ref.~\cite{disturbi},
while discussing the propagation of exponentially localized disturbances in 
the original reference frame. Moreover, it has been shown that only the front
corresponding to the smaller value of $\mu$ is stable, except for some cases
where a nonlinear mechanism intervenes dominating the propagation process
\cite{nonlin}.

At $v = c^2/V_*$ the two intersections degenerate into a single tangency
point. This condition defines $V_*$, which can be interpreted as 
the slowest propagation velocity of initially localized disturbances 
\cite{disturbi}.

\begin{figure}
\psfig{figure=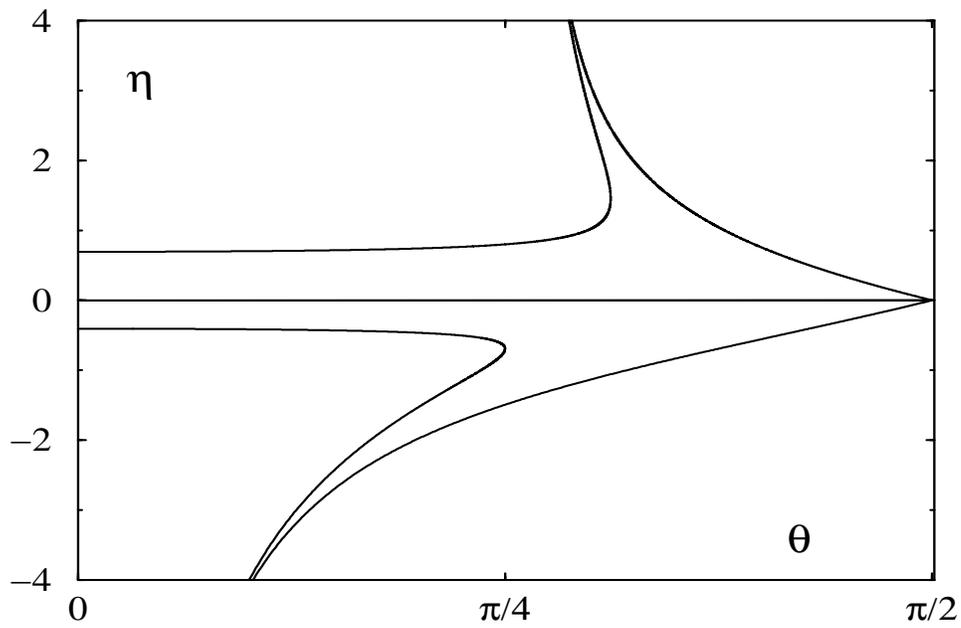,angle=-90,height=10truecm,width=14truecm}
\caption{Boundaries of the STLS versus the tilting 
angle $\vartheta$ for the homogeneous chain ($r=2$, $\varepsilon=1/3$).}
\label{ruotbd}
\end{figure}

The extension of Eq.~(\ref{selfcon}) to the rest of the spectrum requires to
connect $n_\lambda$ and $n_\mu$ with $n_\eta$. In the next section, we will
show how to perform such a step with the help of the entropy potential.
Here, we limit ourselves to discuss the structure of the STLS for different
values of the tilting angle $\vartheta$. In Fig.~\ref{ruotbd}, we report the
borders of the bands, which can be determined from the intersections of
$\cal L$ with the border $\partial {\cal D}$ of the domain of allowed
perturbations (see Fig.~1 of Ref.\cite{lyap1} and Fig.~\ref{bordi}). For
$\vartheta=0$ (temporal case) a single band is present but, as soon as
$\vartheta >0$, a second negative band arises from the intersections with
the branch diverging to $-\infty$ at $\mu=-\mu_c$. For $\vartheta > \pi/4$,
the negative band disappears and a positive band arises from the
intersections with the branch diverging to $+\infty$ with slope $v=c$. A
single band spectrum is again recovered for $\vartheta \ge \vartheta_* =
{\rm atan}(c/V_*)$.

Notice that in symplectic maps, the STLS is symmetric for any value of
$\vartheta$ (see Ref.\cite{lyap1}) so that positive and negative bands
appear and disappear simultaneously.

It is worthwhile to illustrate some of the above considerations in the
simple example of the linear diffusion equation 
\be
\partial_t u = \gamma u + D \partial_x^2 u\quad,
\label{diffusion}
\ee
that arise when dealing with the linear stability analysis of
uniform and stationary solutions $u(x,t)=u^*$ of the scalar version
of Eq.~(\ref{pde}). The expression for $\lambda_{\rm max}(\mu)$ can be 
obtained by assuming $u(x,t)=\exp(\mu x+\lambda t)$, so that
\be
\lambda_{\rm max}(\mu) = \gamma+D\mu^2 \quad .
\ee
Accordingly, Eq.~(\ref{selfcon}) reads as
\be
\label{etamax}
\beta\eta_{\rm max} = \gamma+D\left({v\over c^2}\beta\eta_{\rm max}
\right)^2  \quad .
\ee
On the other hand, the model equation in the rotated frame can be obtained
from the substitutions
\bey
\label{trasdif}
&&\partial_t \to  \beta \left(\partial_{t'} -v\partial_{x'} \right) \\ 
&&\partial_x \to \beta \left( {v\over c^2}\partial_{t'} + \partial_{x'}
\right) \quad . \nonumber
\eey
By introducing the usual Ansatz for the shape of the perturbation,
\be
\label{uprimed}
u(x',t') \sim  \exp \left[i\kappa x' + (\eta+i\Omega)t'\right] \quad ,
\ee
separating the real from the imaginary part, and eliminating
$\Omega$, we obtain the integrated density of spatiotemporal exponents
\be
\label{rls}
\kappa(\eta,v) = \beta\left( 1-2D{v^2\over c^4}\beta\eta\right)
\sqrt{{v^2\over c^4}(\beta\eta)^2 - { \beta\eta \over D}
+ {\gamma\over D}} 
\ee
and the corresponding STLS $\eta(\kappa,v)$. Dimensional analysis shows that
$\kappa$ is an inverse length, as expected for a density of exponents.
Notice that Eq.~(\ref{etamax}) is recovered, by setting $\kappa = \Omega = 0$
in Eq.~(\ref{rls}). 

In this and more general continuous models, we should remark that the line
$\cal L$ intersects $\lambda_{\rm max}(\mu)$ twice for any arbitrarily
small $v$. This is because the Laplacian operator sets no upper limit to the 
propagation velocity of disturbances.

\section{From the entropy potential to dynamical invariants}

The present section is devoted to establish the consequences of the
existence of the entropy potential on the Lyapunov spectra and other
dynamical indicators such as the Kolmogorov-Sinai entropy and the
Kaplan-Yorke dimension of the attractor. In order to keep the notations as
simple as possible, we assume that time and space coordinates are scaled in
such a way that $c=1$.

\subsection{Spatiotemporal exponents} 
 
\begin{figure}
\psfig{figure=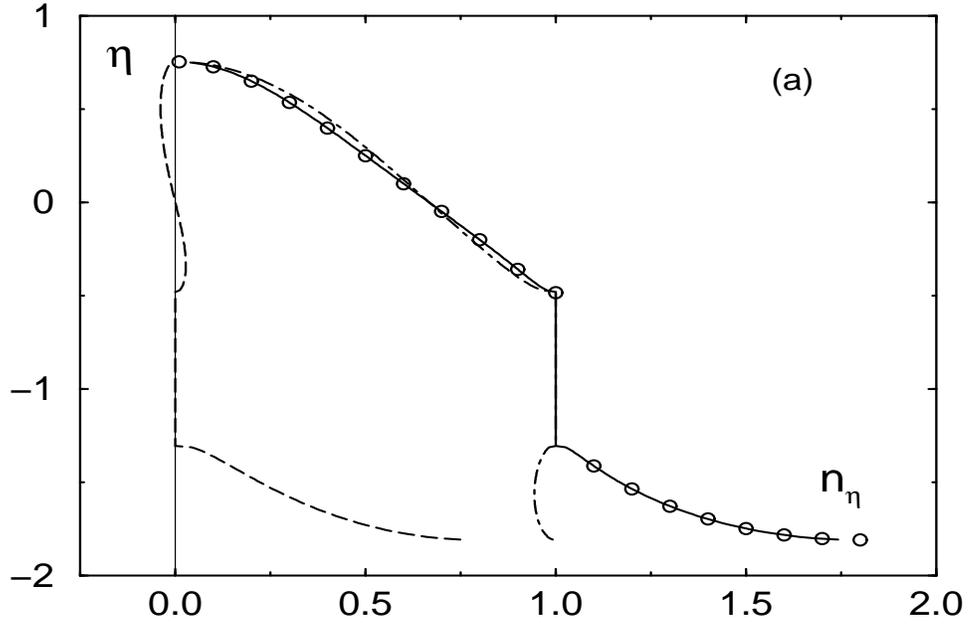,angle=-90,height=10truecm,width=14truecm}
\psfig{figure=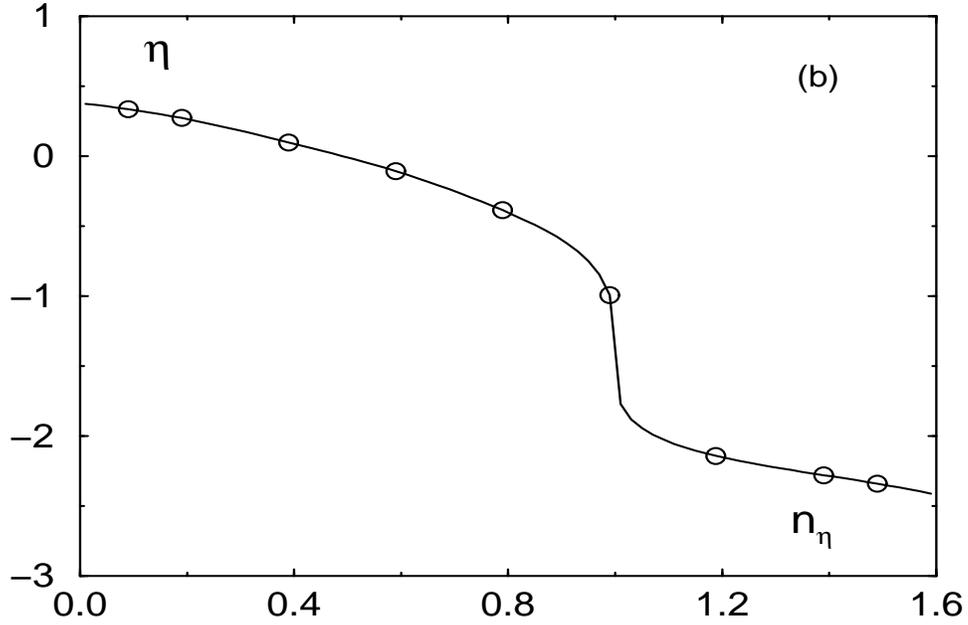,angle=-90,height=10truecm,width=14truecm}
\caption{Comparison between the STLS obtained by direct
numerical computation and formula (\ref{formula}) for (a) homogeneous 
($r=2$) with $v=4/5$ and (b) logistic CMLs with $v=3/5$ (in both 
cases $\varepsilon=1/3$).}
\label{spetr}
\end{figure}

The very existence of the entropy potential $\Phi$ implies that the Lyapunov
spectrum in a frame tilted at an angle $\vartheta$ (recall that $\vartheta$ is
the angle from the $\lambda$-axis) can be obtained by computing the
derivative of $\Phi$ along the direction
$\vec u = (\sin \vartheta, \cos \vartheta)$ in the $(\mu,\lambda)$ plane. In
fact, this is a straightforward generalization of the previous findings that
$n_\mu$ and $n_\lambda$ are the derivatives of $\Phi$ along the $\mu$ and
$\lambda$ direction, respectively. Accordingly, the STLS is linked to the
TLS and SLS by the following general equation
\be
\label{formula}
n_\eta(v,\eta) = \vec u \cdot \nabla \Phi =
\beta \left [ v n_\mu + n_\lambda \right ] \quad ,
\ee
where $\nabla=(\partial_\mu,\partial_\lambda)$ is the gradient in the
$(\mu,\lambda)$ plane, and the r.h.s of the above formula is evaluated for
\bey
&&\mu = v \beta\eta \\
&&\lambda = \beta\eta \quad .
\eey
Such a relation can be directly verified for the diffusion equation from
Eqs.~(8) of Ref.\cite{lyap2} and Eq.~(\ref{rls}). Further, more significative 
tests have been performed by checking numerically the validity of 
Eq.~(\ref{formula}) in some lattice models involving, e.g., logistic and 
homogeneous chains, the spectra of which are reported in see Fig.~\ref{spetr} 
(notice that, since the time units have not been renormalized in the tilted 
frame, the factor $\beta$ is no longer needed).

\subsection{Entropy}

The connection between the chronotopic approach and the Kolmogorov-Sinai
entropy $H_{KS}$ was already discussed in \cite{lyap2}. Here we recall
the main concepts both for the sake of completeness and since it can be
more effectively phrased with the help of the spatio-temporal
representation.

As usual, we shall refer to Pesin formula \cite{eck} as a way to estimate
the rate of information-production from the positive Lyapunov exponents.
Although it provides an upper bound, there is numerical evidence that the
bound is actually saturated in generic models. The reference to Pesin's
formula suggests that $H_{KS}$ is an extensive quantity \cite{grass}.
For this reason, it is convenient to introduce the entropy density
$h_\lambda$ which, in the thermodynamic limit, can be obtained from the
integral of the positive part of the Lyapunov spectrum.

It is natural to extend the definition of entropy to a generic reference
frame,
\be
\label{pesin}
h_\eta = \int_0^{\eta_{\rm max}} n_\eta (\vartheta,\eta) d\eta
\quad ,
\ee
where the integral is performed along the line $\cal L$ and 
$\eta_{\rm max}$ is the maximum value of $\eta$ which is reached in the
intersection point between $\cal L$ and $\cal D$.
In the limit $\vartheta = 0$, the above equation reduces to the usual 
definition of $h_\lambda$, while for $\vartheta = \pi/2$ it reduces to
the ``spatial'' entropy $h_\mu$.

The existence of an entropy potential implies that
\be
h_\eta(\vartheta) = \cases{ h_\lambda & $ |\vartheta| < \vartheta_*$ \cr
	                   h_\mu     & $ |\vartheta|\ge \vartheta_*$}
\label{cases}
\ee
where $\vartheta_*$ is the value for which the line $\cal L$ is tangent to
$\cal D$. This is immediately seen by combining the observation that
$h_\eta$ follows from an integral along the straight line defining the
corresponding $\lambda$-axis with the observation that one of the two
extrema is always the same (the origin) while the others lie along an
equipotential line.

The independency of $h_\eta$ of $v$ has an immediate physical
interpretation, which we also consider as a strong argument in favour of
the existence of an entropy potential. The Kolmogorov-Sinai entropy density
is, in fact, the amount of information necessary to characterize a
space-time pattern of temporal duration $T$ and spatial extention $S$ (apart
from the information flow through the boundaries \cite{grass}) divided by
its area $S \times T$. Therefore, $h_{KS}$ is expected to be independent of
the way the axes are oriented in the plane, i.e. of the velocity $v$. As a
consequence, $h_\eta = h_{KS}$ for all $v<1$.

The above conclusion still holds when the STLS exhibits a positive band as
well (which is always the case in continuous models), provided that the
content of such a band is discarded. Accordingly, we can conclude that the
new degrees of freedom, associated to the positive band, which appear in the
rotated frame are just physically irrelevant directions which turn the
original attractor into a repellor. If $v> c^2/V_*$, the two bands merge
together and it is not anymore possible to distinguish between unstable but
irrelevant directions and the unstable manifold of the original attractor.
Presumably, this means that the repellor is turned into a strange repellor
with a singular measure along some (all) unstable directions.

\subsection{Dimension}

A second important indicator of the ``complexity'' of a spatiotemporal dynamics
is the fractal dimension of the underlying measure. An upper bound $D_{KY}$ to 
it is given by the Kaplan-Yorke formula \cite{eck}. The existence of a limit 
Lyapunov spectrum, implies that $D_{KY}$ is proportional to the system size,
so that it is convenient to introduce the dimension density 
$d_{KY}$\cite{grass}, i.e. the number of independent degrees of freedom
actively involved in the asymptotic evolution per unit length.
In the framework of the present paper, it is natural to extend the concept
of dimension density to a frame oriented in a generic way in the space-time 
plane. A straightforward generalization of the Kaplan-Yorke formula leads to 
the integral equation,
\be
\label{ky}
\int _0^{d_{KY}} \eta(v,n_\eta) dn_\eta = 0 \quad .
\ee
As for the entropy density, Eq.~(\ref{ky}) can be more easily interpreted with
reference to the $(n_\mu,n_\lambda)$ plane. In this representation the 
entropy potential is
\be
\label{conphi}
\tilde \Phi = \lambda n_\lambda + \mu n_\mu - \Phi \quad ,
\ee
In fact, the curve implicitely defined by the constraint (\ref{ky}) is the 
equipotential line $\cal C$ 
\be
  \tilde \Phi (n_\mu,n_\lambda) = 0  \quad .
\ee
The dimension density $d_{KY}(v)$ can, in turn, be determined from 
Eq.~(\ref{formula}) at the intersection point between $\cal C$ and the
image of $\cal L$ in the plane $(n_\mu,n_\lambda)$. 

\begin{figure}
\psfig{figure=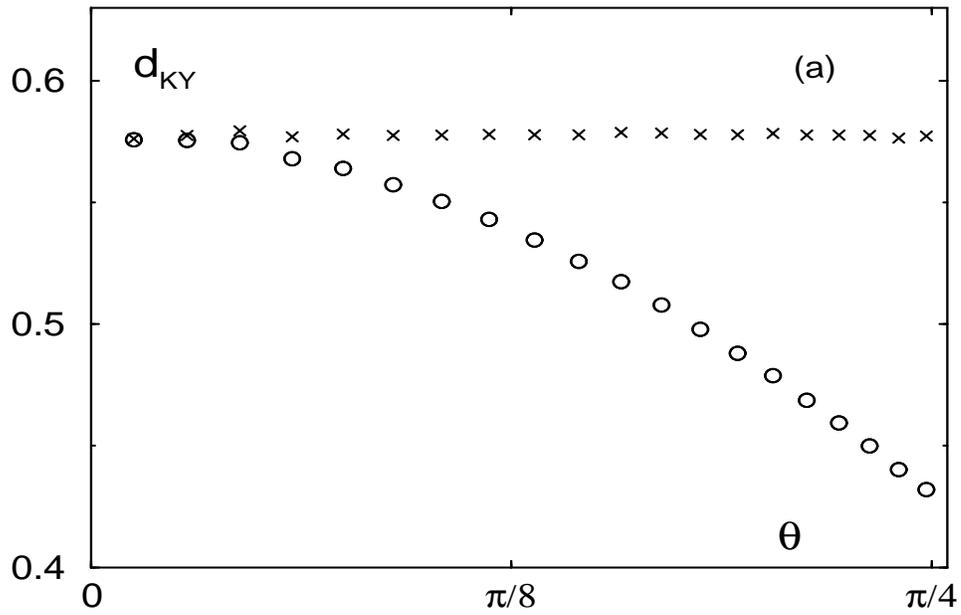,angle=-90,height=10truecm,width=14truecm}
\psfig{figure=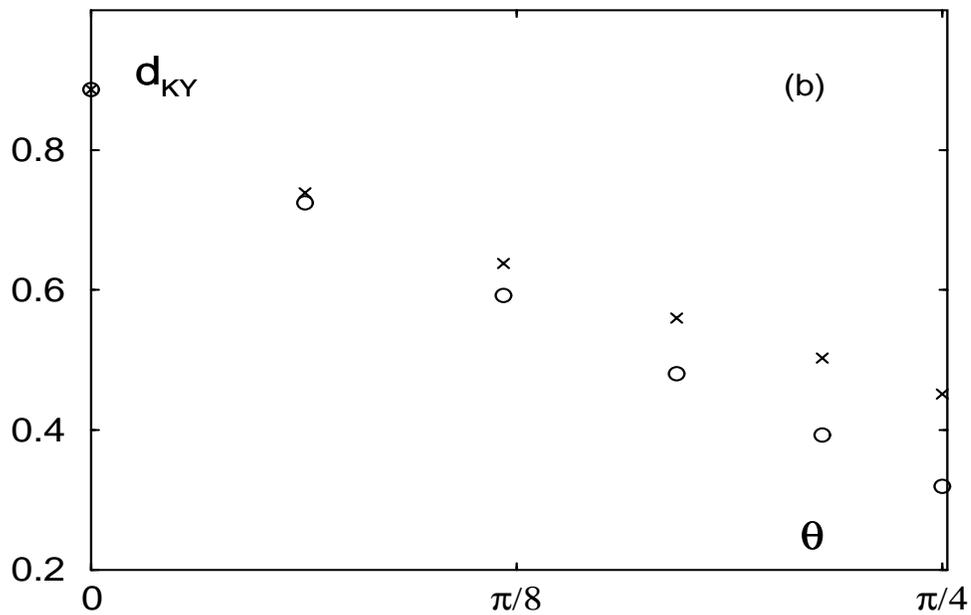,angle=-90,height=10truecm,width=14truecm}
\caption{Kaplan-Yorke dimension density $d_{KY}$ obtained from 
the STLS versus the tilting angle $\vartheta$ for (a) homogeneous
($r=1.2$) and (b) logistic CMLs: in both cases 
$\varepsilon=1/3$. Circles refer to the scaling factor $c=1$, while
crosses correspond to $c=3$, $+\infty$ in (a), (b), respectively. 
}
\label{dim}
\end{figure}

At variance with the entropy density, $d_{KY}(v)$ changes with $v$ (see 
Fig.~\ref{dim}) even if we avoid considering the second positive band.
In fact, while $h_\eta$ is an information divided by a space-time area, 
$d_{KY}(v)$ is a number of degrees of freedom divided by a length, measured 
orthogonally to the propagation axis. Thus, at least from a dimensional 
point of view, it is meaningless to compare $d_{KY}(v)$ for different 
velocities. However, one can reduce temporal to spatial lengths by
introducing the scaling factor $c$ and, in turn, ask himself how the
dimension changes with $c$. It is easily seen that the scaling dependence
on $c$ is expressed by the following relation,
\be
\label{scale}
d_{KY}(v,c_1) \sqrt{1 + \left ({ v \over c_1 } \right )^2} = 
d_{KY}(v,c_2) \sqrt{1 + \left ({ v \over c_2 } \right )^2} \quad , 
\ee
The (completely arbitrary) choice of $c$ reflects in different dependences of 
$d_{KY}$ on $v$. A natural procedure to fix $c$ is by minimizing the 
dependence of $d_{KY}$ on the observation angle. This amounts to choosing the 
time units in such a way as to make the 2D pattern as isotropic as 
possible. In homogeneous CMLs, the procedure is so effective that a suitable 
choice of $c$ allows removing almost completely the velocity dependence as 
seen in Fig.~\ref{dim}a, where the results for the natural value $c=1$ are 
compared with those for $c=3$.\par
More in general, however, it is not possible to achieve such a complete
success. This is, for instance, the case of the logistic CML, where the 
dimension drop for $c=1$ is too large to be compensated by any choice of $c$ 
(see Fig.~\ref{dim}b, where the curve for $c=1$ is compared with the best
results obtained for $c=+\infty$).

A further indicator which is sometimes useful in characterizing the chaoticity
of a given extended system is the dimension density $d_u$ of the unstable 
manifold. This dimension is nothing but $n_\eta$ in the point where $\eta=0$,
i.e. in the origin, and its expression simply reads as
\be
d_u = \beta n_\lambda(0,0) \quad .
\ee
being $n_\mu(0,0) \equiv 0$. The choice $c=+\infty$ of the scaling factor 
removes exactly the dependence on the orientation of the reference frame.
This choice is equivalent to measuring lengths in the untilted frame.

\section{Conclusions and open problems}

In this paper we have provided further numerical evidence for the existence
of an entropy potential in 1D spatially extended dynamical systems. The
result appears to follow from the possibility to order the Lyapunov vectors
according to their average wavenumber that thus becomes equivalent to
the integrated density of exponents. This interpretation, if confirmed,
would imply the possibility to define and attribute some meaning to
the imaginary part of the expansion rates. This line of thought is very
reminiscent of the rotation numbers introduced by Ruelle in the context
of Hamiltonian systems \cite{rue}. It is certainly worth to fully explore
this route in the hope to arrive at a more rigorous justification of our
theoretical construction : work in this direction is in progress.

Furthermore, we have seen that, among the consequences of the existence of an
entropy potential, there is the possibility to insert very coherently in
this scheme a more general class of Lyapunov exponents corresponding to
various directions of propagation in the space-time. As a result, we
have found that the asymmetry between the temporal and the spatial axes
can be revealed by the dependence of the density of dimension on the
orientation of the reference frame. This is not true for the
Kolmogorov-Sinai entropy density which, as already remarked in \cite{lyap2},
should be regarded as a ``super-invariant'' dynamical indicator.

\acknowledgments

We thank P. Grassberger, H. Kantz, and A. Pikovsky for useful 
discussions. We also acknowledge the hospitality of ISI-Torino during the
activity of the EC Network CHRX-CT94-0546, and one of us (A.T.) also
the european contract ERBCHRX-CT94-0460.

\begin{table}
\vskip 1 truecm
\begin{tabular}{ccccc}
\multicolumn{1}{c}{Path}
& \multicolumn{1}{c}{$\Delta t=0.2$}
& \multicolumn{1}{c}{$\Delta t=0.1$}
& \multicolumn{1}{c}{$\Delta t=0.05$} \\ 
\hline
\hline

$(0,0)\to(0,1)\to(2,1)$ &  -1.1419  & -1.1868  &  -1.2303  \\               
$(0,0)\to(2,0)\to(2,1)$ &   -1.1358  &   -1.1838  &    -1.2236  \\
       &            &            &             \\
Relative difference&   0.5 \% &  0.2 \% & 0.5 \% \\
\end{tabular} 
\vskip 1 truecm
\caption[tabone]{
Entropy potential $\Phi$ for model (\ref{rndosci}) ($S_0=1/3$, $\sigma=0.1$,
$\omega_0=1$, $D=1$) computed by integrating along two different piecewise 
linear 
paths of the $(\mu,\lambda)$ plane, and for three different time steps. In the 
last row the relative difference between the two values is reported: 
relative statistical error in the computation of Lyapunov exponents are 
of the order of $10^{-3}$.
}
\label{tab1}
\end{table}
\end{document}